\documentclass{PoS}
\usepackage{amsmath}

\title{ Three-quark systems in MA and MC projected QCD}

\ShortTitle{Three-quark systems in MA and MC projected QCD }

\author{\speaker{Hideaki Iida}%
\\
       Yukawa Institute for Theoretical Physics, Kyoto University \\
Kitashirakawaoiwake, Sakyo, Kyoto 606-8502, Japan\\
       E-mail: \email{iida@yukawa.kyoto-u.ac.jp}}
\author{Naoyuki Sakumichi and Hideo Suganuma\\
 Department of Physics, Kyoto University, \\
Kitashirakawaoiwake, Sakyo, Kyoto 606-8502, Japan\\
         E-mail: \email{sakumichi@scphys.kyoto-u.ac.jp}, 
\email{suganuma@scphys.kyoto-u.ac.jp}
}

\abstract{
We study three quark systems in Maximally Abelian (MA) and 
Maximal Center (MC) projected QCD on quenched SU(3) lattice, 
and also in the monopole/photon part, 
where only the color-electric/magnetic current exists, 
using the Hodge decomposition.
First, we perform the quantitative study of 
the three-quark (3Q) potential $V_{\rm 3Q}$ and 
the string tension $\sigma_{\rm 3Q}$ in baryons. 
For MA projected QCD, the monopole part and MC projected QCD, 
we find that the confinement potential in $V_{\rm 3Q}$ obeys the Y-Ansatz 
and the string tension $\sigma_{\rm 3Q}$ is 
approximately equal to that in SU(3) QCD.
The universality of the string tension, 
$\sigma_{\rm 3Q} \simeq \sigma_{\rm Q\bar Q}$, is also found 
between the 3Q and the ${\rm Q\bar Q}$ potentials.
We find a strong similarity of the inter-quark potential 
between the monopole part and MC projected QCD. 
In contrast, almost no confinement force is found in the inter-quark potential
in the photon part.
Next, we study the spectrum of light hadrons in 
MA projected QCD and the monopole/photon part, 
paying attention to the N-$\Delta$ mass splitting. 
We find that the N-$\Delta$ mass splitting is significantly reduced in 
MA projected QCD and the monopole part, 
where the one-gluon-exchange effect or 
the Coulomb-potential part is largely reduced 
due to the Abelianization or the Hodge decomposition.
This fact seems to indicate that 
the main origin of the mass splitting is one-gluon exchange.}

\FullConference{The XXVI International Symposium on Lattice Field Theory \\
                July 14 - 19, 2008\\
                Williamsburg, Virginia, USA}

\begin{document}

\section{Introduction}
Color confinement in QCD is one of the most challenging 
problems in physics. 
One of the possible scenarios of color confinement is the 
dual-superconductor picture, proposed by Nambu \cite{N74}, 't~Hooft and 
Mandelstam in 70's. 
The key point of color confinement is one-dimensional squeezing of the 
color-electric flux, and the dual-superconductor picture 
realizes such squeezing due to the dual Meissner effect, 
which is the dual version of the Meissner effect in superconductors. 

However, there are two large gaps between 
the dual-superconductor picture and QCD. 
One gap is the non-Abelian nature of QCD. 
While QCD is a non-Abelian gauge theory, 
the dual-superconductor picture is based on the Abelian gauge theory 
subject to the Maxwell-type equation, 
where electro-magnetic duality is manifest.
The other gap is the existence of the monopole degrees of freedom. 
The dual-superconductor picture requires condensation of color-magnetic 
monopoles as the key concept, while QCD does not have such 
a monopole as the elementary degrees of freedom. 
To compensate these two gaps, 
Maximally Abelian (MA) gauge is a suitable gauge choice \cite{tH81}. 
The important point is that the off-diagonal gluons 
have a large effective mass of about 1GeV in MA gauge \cite{AS99}. 
Therefore, only the diagonal gluons are relevant 
in infrared region in MA gauge. 
The other important point is that color-magnetic monopoles appear 
as topological objects reflecting the nontrivial homotopy group: 
$\Pi_2({\rm SU}(N_c)/{\rm U(1)}^{N_c-1})={\bf Z}_\infty^{N_c-1}$.
Therefore, QCD is reduced to an Abelian theory with monopoles 
in MA gauge in infrared region. 
In the following, we call the Abelian theory as ``MA projected QCD''. 

MA projected QCD is decomposed into two parts by the Hodge decomposition. 
One is ``monopole part'', where the color-magnetic monopole current $k_\mu$ 
exists without the color-electric current $j_\mu$. 
The other is ``photon part'', where the color-electric current $j_\mu$ 
exists without the color-magnetic current $k_\mu$. 
Since monopoles are considered to contribute the nonperturbative phenomena 
such as confinement, roughly speaking, the monopole part is responsible 
for ``nonperturbative'' phenomena related to the vacuum structure of 
QCD, and the photon part is responsible for ``perturbative'' phenomena. 
Actually, the ${\rm Q\bar Q}$ potential can be well separated 
into the linear confinement potential in the monopole part and 
the Coulomb potential in the photon part by the Hodge decomposition. 
Moreover, the chiral symmetry is spontaneously broken 
and instantons survive in the monopole part, while 
no spontaneous chiral symmetry breaking appears and 
no instanton survives in the photon part \cite{M95,SM96}. 
Then, using the monopole/photon decomposition, 
we can clarify whether each QCD phenomenon mainly originates 
from ``nonperturbative'' nature or ``perturbative'' nature. 

Similar essence of color confinement can be also extracted in 
Maximal Center (MC) gauge \cite{DFGO97}, which is a gauge choice that 
extracts the degrees of freedom of the center group ${\rm Z}_3$ of QCD. 
A four-dimensional ${\rm Z}_3$-spin system can be obtained 
through MC projection after MC gauge fixing, and 
the linear confinement potential can be well extracted as 
the ${\rm Q\bar Q}$ potential in this system \cite{DFGO97}. 

Recently, the detailed analysis of the three-quark (3Q) potential 
was performed in lattice QCD \cite{TS0102}, 
which reveals the Y-type linear confinement \cite{TS0102} 
and the Y-shaped flux-tube formation in baryons \cite{I0304}.
In this paper, aiming to clarify the quark confinement mechanism in baryons, 
we perform the quantitative lattice study of the 3Q potential 
in MA projected QCD, the monopole/photon part \cite{SYSTHO07} and 
MC projected QCD.
(The semi-quantitative study of the 3Q potential was done 
in MA gauge in Ref.\cite{I0304}.)
We also study the spectrum of light hadrons, i.e., N, 
$\Delta$, $\pi$ and $\rho$, in MA projected QCD and the monopole/photon part, 
paying attention to the N-$\Delta$ mass splitting.

\section{MA gauge fixing, MA projection, Hodge decomposition and MC gauge fixing}
In lattice QCD, Maximally Abelian (MA) gauge is defined 
so as to maximize the quantity 
$R_{\rm MA} [U_\mu(s)]\equiv {\rm Re}
\sum_{s,\mu}{\rm Tr}\left(U_\mu(s)\vec H U_\mu^\dagger(s) \vec H\right)$ 
by SU(3) gauge transformation, 
with the link-variable $U_\mu(s)$ and the Cartan subalgebra $\vec H$ of SU(3). 
We denote the MA gauge fixed configuration by $U_\mu^{\rm MA}(s) 
\in {\rm SU(3)}_c$. After the MA gauge fixing, 
we extract Abelian link-variable $u_\mu^{\rm Abel}(s) 
\in {\rm U(1)}^2 \subset {\rm SU(3)}_c$ 
from $U_\mu^{\rm MA}(s) \in {\rm SU(3)}_c$ by maximizing 
${\rm Re} {\rm Tr} \left( u_\mu^{\rm Abel}(s)U_\mu^{\rm MA}(s)^\dagger\right)$.
MA projection is defined by the replacement of 
the SU(3) link-variable $\{U_\mu^{\rm MA}(s)\}$ by the Abelian link-variable 
$\{u_\mu^{\rm Abel}(s)\}$.

In the MA projected QCD, there appear not only the electric current $j_\mu(s)$ 
but also the magnetic (monopole) current $k_\mu(s)$. 
By the Hodge decomposition, as will be explained below, 
the MA projected QCD can be decomposed into the ``monopole part'' 
which only includes the magnetic current $k_\mu$ and 
the ``photon part'' which only includes the electric current $j_\mu$. 
We define the gauge field $\theta^i_\mu(s)\in (-\pi,\pi]$ ($i$=1,2,3) as 
$u_\mu^{\rm Abel} (s)=
{\rm diag}(e^{i\theta^1_\mu}(s), e^{i\theta^2_\mu}(s), e^{i\theta^3_\mu}(s))$. 
The field strength tensor $\theta_{\mu\nu}^i(s) \in (-\pi,\pi]$ 
is defined as $(\partial\land\theta)_{\mu\nu}^i
\equiv \partial_\mu\theta_\nu^i-\partial_\nu\theta_\mu^i
=\theta_{\mu\nu}^i+2\pi n_{\mu\nu}^i$, with $n_{\mu\nu}^i \in {\bf Z}$.
From $\theta_{\mu\nu}^i(s)$, 
the electric current $j_\nu^i(s)$ and 
the magnetic current $k_\nu^i(s)$ are defined as 
$j_\nu^i \equiv \partial_\mu \theta_{\mu\nu}^i$ and 
$k_\nu^i \equiv \partial^*_\mu\theta_{\mu\nu}^i
=-2\pi\partial_\mu^*n_{\mu\nu}^i$, with 
$^*\theta_{\mu\nu}^i\equiv \epsilon_{\mu\nu\rho\sigma}\theta_{\rho\sigma}^i/2$.
For simple notation, we hereafter abbreviate the index $i$.
Since the photon part has only the electric current by definition, 
the field strength $\theta_{\mu\nu}^{\rm Ph}$ in the photon part obeys 
$j_\nu=\partial_\mu\theta_{\mu\nu}^{\rm Ph}$ and 
$\partial^*_\mu\theta_{\mu\nu}^{\rm Ph}=0$. 
Then, the gauge field $\theta_\mu^{\rm Ph}$ in the photon part obeys 
the four-dimensional Poisson equation 
$j_\nu(s)=\partial^2\theta_\nu^{\rm Ph}(s)$, 
by taking U(1) Landau gauge fixing, $\partial_\mu\theta_\mu^{\rm Ph}(s)=0$,
for the residual U(1)$^2$ gauge symmetry.
Solving the Poisson equation, $\theta_\mu^{\rm Ph}(s)$ can be obtained. 
The gauge field $\theta^{\rm Mo}(s)$ in the monopole part 
is obtained as 
$\theta_\mu^{\rm Mo}(s)=\theta_\mu(s)-\theta^{\rm Ph}_\mu(s)$. 

Maximal Center (MC) gauge fixing \cite{DFGO97} is defined to maximize 
$\sum_{s, \mu}|{\rm Tr} \ U_\mu(s)|^2$ by gauge transformation, 
and the link-variable maximally approaches 
to ${\rm Z}_3$ element in MC gauge.
After the MC gauge fixing, ${\rm Z}_3$ link-variable $z_\mu(s) 
\in {\rm Z}_3$ is extracted by maximizing 
${\rm Re} {\rm Tr} \left( z_\mu(s)U_\mu(s)^\dagger\right)$, 
and MC projection is defined as the replacement 
by the ${\rm Z}_3$ link-variable 
$z_\mu(s)=e^{2\pi i m(s)/3}\cdot {\bf 1} \ (m(s)=0, 1, 2)$.

\section{The three-quark potential in MA projected QCD}
The three-quark (3Q) potential $V_{\rm 3Q}$ \cite{TS0102} is 
obtained as 
$V_{\rm 3Q}=-\lim_{T\rightarrow\infty}\frac{1}{T}{\rm ln} 
\langle W_{\rm 3Q}(T)\rangle$ 
from the 3Q Wilson loop 
$W_{\rm 3Q}(T)\equiv 
\frac{1}{3!}\epsilon_{abc}\epsilon_{a^\prime b^\prime c^\prime} 
U_1^{aa^\prime}U_2^{bb^\prime}U_3^{cc^\prime}$ 
with 
$U_k\equiv P \exp \left\{ ig\int_{\Gamma_k}dx_\mu A^\mu(x)\right\}$ 
$(k=1,2,3)$ as the path-ordered product along $\Gamma_k$ as shown in Fig.~1. 
We calculate the 3Q potential for 120 different patterns of 3Q systems 
in lattice QCD with $16^3\times 32$ and 
$\beta\equiv 6/g^2=6.0$, i.e., $a \simeq 0.1{\rm fm}$ for the lattice spacing.
To enhance the ground-state component, we adopt the smearing method. 
For the calculation, we use 100 gauge configurations generated with NEC SX-8R 
in Osaka University. 

We thus obtain the 3Q potential $V_{\rm 3Q}$ in SU(3) QCD, 
MA projected QCD, the monopole/photon part and MC projected QCD, respectively. 
For the quantitative analysis, 
we consider the functional fit for the obtained lattice data 
of the 3Q potential in each part using the Y-Ansatz \cite{TS0102}, 
\begin{eqnarray}
V_{\rm 3Q}=-A_{\rm 3Q}\sum_{i<j}\frac{1}{|{\bf r}_i-{\bf r}_j|}
+\sigma_{\rm 3Q}L_{\rm min}+C_{\rm 3Q}
=-A_{\rm 3Q}/L_{\rm Coul}+\sigma_{\rm 3Q}L_{\rm min}+C_{\rm 3Q},
\label{cornell}
\end{eqnarray}
with the static quark location ${\bf r}_i$ ($i$=1,2,3), 
$L_{\rm Coul}\equiv (\sum_{i<j}\frac{1}{|{\bf r}_i-{\bf r}_j|})^{-1}
=(1/a+1/b+1/c)^{-1}$ and $L_{\rm min}\equiv {\rm AP}+{\rm BP}+{\rm CP}$, 
the minimum length connecting the three quarks, as shown in Fig.~2. 
The 3Q potential in SU(3) QCD is known to be well fit by the Y-Ansatz 
within 1\%-level deviation \cite{TS0102}. 

\begin{figure}
\begin{minipage}{.45\linewidth}
\centering
\includegraphics[width=2.8cm]{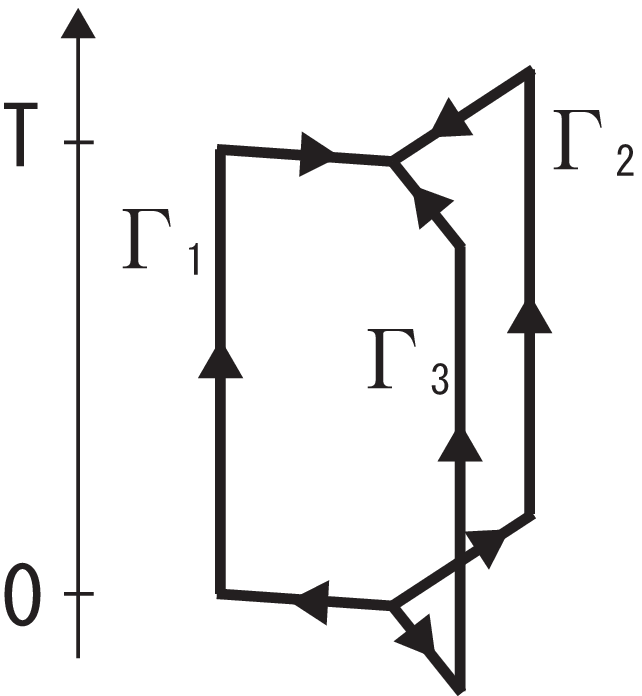}
\caption{The schematic description of 
the three-quark Wilson loop $W_{\rm 3Q}$.}
\label{Wlp}
\end{minipage}
\hspace{0.5cm}
\begin{minipage}{.45\linewidth}
\centering
\includegraphics[width=4.1cm]{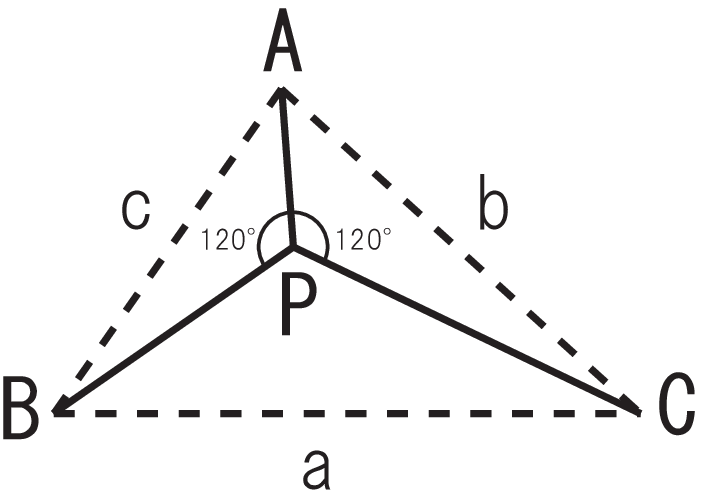}
\caption{The system of static three quarks located at A, B and C. 
P denotes the Fermat point.}
\end{minipage}
\end{figure}

\section{Numerical results of the three-quark potential in MA and MC projected QCD}
Now, we show the numerical results of the three-quark (3Q) potential.
Figure 3 and 4 show the 3Q potential in SU(3) QCD, MA projected QCD, 
the monopole part and MC projected QCD, plotted against $L_{\rm min}$. 
For MA projected QCD, the monopole part and MC projected QCD, 
we find that the 3Q potential $V_{\rm 3Q}$ is approximately 
a single-valued function of $L_{\rm min}$, although $V_{\rm 3Q}$ 
generally depends on three independent variables, e.g., $a$, $b$ and $c$.
We note that the results in the monopole part and MC projected QCD 
are very similar apart from an irrelevant constant. 
Figure 5 and 6 show the 3Q potential in the photon part 
plotted against $L_{\rm min}$ and $L_{\rm Coul}$, respectively.
The 3Q potential in the photon part is not a single-valued function 
of $L_{\rm min}$, but approximately a single-valued function of $L_{\rm Coul}$ 
except for $L_{\rm Coul} \le 1$, where the discretization error is large. 

\begin{figure}[h]
\begin{minipage}{.45\linewidth}
\centering
\includegraphics[width=5.8cm,clip]{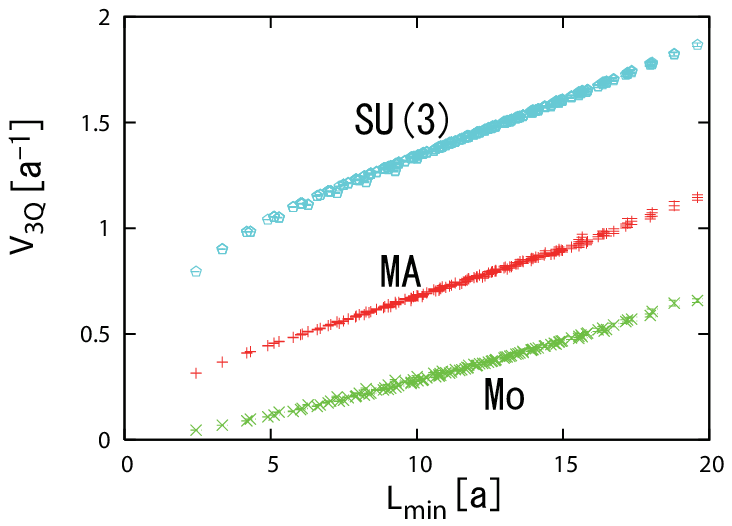}
\vspace{-0.3cm}
\caption{The 3Q potential $V_{\rm 3Q}$ in SU(3) QCD, MA projected QCD (MA) and 
the monopole part (Mo) plotted against $L_{\rm min}$. }
\label{Wlp}
\end{minipage}
\hspace{0.5cm}
\begin{minipage}{.45\linewidth}
\centering
\includegraphics[width=5.8cm,clip]{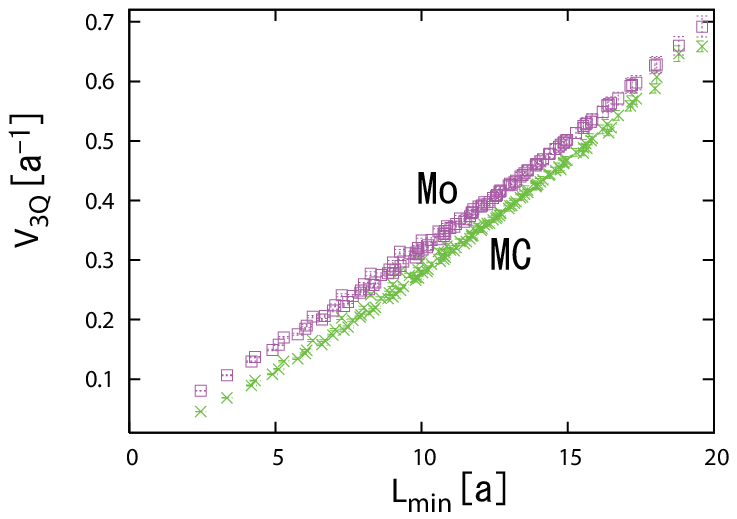}
\vspace{-0.3cm}
\caption{The 3Q potential $V_{\rm 3Q}$ 
in the monopole part (Mo) and MC projected QCD (MC) 
plotted against $L_{\rm min}$. }
\label{Wlp}
\end{minipage}
\vspace{-0.35cm}
\end{figure}

\begin{figure}[hb]
\begin{minipage}{.45\linewidth}
\centering
\includegraphics[width=5.8cm]{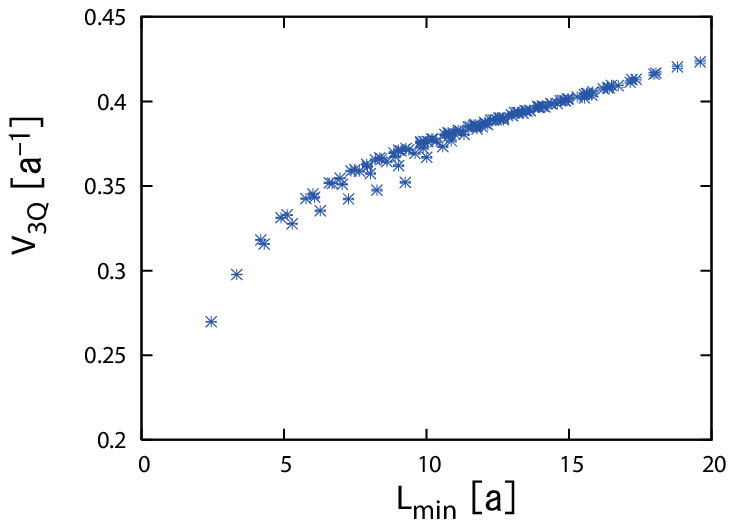}
\vspace{-0.3cm}
\caption{The 3Q potential $V_{\rm 3Q}$ 
in the photon part plotted against $L_{\rm min}$.}
\label{u1ph_lmin}
\end{minipage}
\hspace{0.5cm}
\begin{minipage}{.45\linewidth}
\centering
\includegraphics[width=5.8cm]{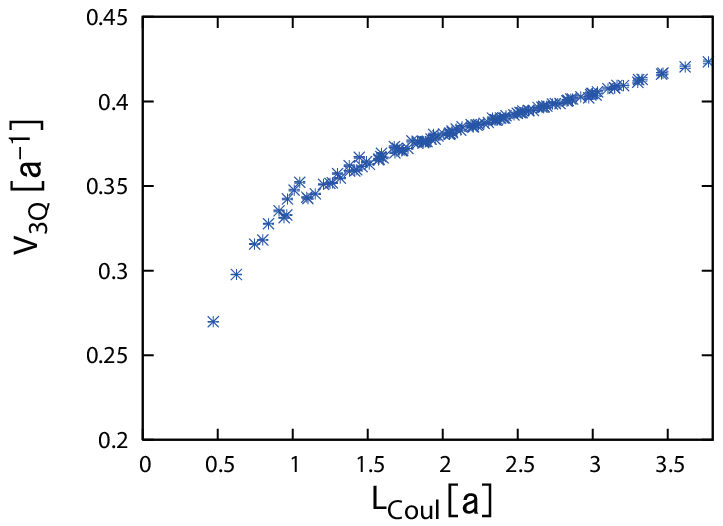}
\vspace{-0.3cm}
\caption{The 3Q potential $V_{\rm 3Q}$ 
in the photon part plotted against $L_{\rm Coul}$.}
\label{u1ph_Coul}
\end{minipage}
\end{figure}

We next do the fit analysis 
by the Y-Ansatz, i.e., the functional form of Eq.~(\ref{cornell}). 
For the monopole part and MC projected QCD, 
short-distance lattice data are excluded in the fit analysis.
For MA projected QCD, the monopole part and MC projected QCD, 
we find that the 3Q potential $V_{\rm 3Q}$ obeys the Y-Ansatz 
for the confinement potential and includes almost no Coulomb potential, 
which makes $V_{\rm 3Q}$ be a single-valued function of $L_{\rm min}$.
The reduction of the Coulomb potential in MA projected QCD 
can be explained by the reduction of the gluon components from 8 to 2 
through the Abelianization.
On the other hand, the 3Q potential $V_{\rm 3Q}$ in the photon part includes 
almost no linear potential proportional to $L_{\rm min}$, 
which makes $V_{\rm 3Q}$ be a single-valued function of $L_{\rm Coul}$.

We summarize in Table 1 the string tension $\sigma_{\rm 3Q}$ 
in the 3Q potential and 
the string tension $\sigma_{\rm Q\bar Q}$ in the ${\rm Q\bar Q}$ potential 
for SU(3) QCD, MA projected QCD, the monopole part, 
MC projected QCD and the photon part.
We find the approximate equality of the string tension among 
SU(3) QCD, MA projected QCD, the monopole part and MC projected QCD, 
as well as the universality of the string tension as 
$\sigma_{\rm 3Q}\simeq \sigma_{\rm Q\bar Q}$.
In contrast, almost zero string tension emerges in the photon part.

\begin{table}[h]
\label{tab1}
\begin{center}
\begin{tabular}{cccccc}
\hline
\hline
& SU(3) QCD& MA projected QCD& Monopole part& MC projected QCD & Photon part\\
\hline
$\sigma_{\rm 3Q}$ 
& 0.046 & 0.0456 & 0.0382 & 0.0372 & 0.0021 ($\sim$ 0) \\
$\sigma_{\rm Q\bar Q}$ 
& 0.0506 & 0.0439 & 0.0402 & 0.0361 & 0.0041 ($\sim$ 0) \\
\hline
\hline
\end{tabular}
\caption{The string tension $\sigma_{\rm 3Q}$ in the 3Q potential and 
the string tension $\sigma_{\rm Q\bar Q}$ in the ${\rm Q\bar Q}$ potential 
in SU(3) QCD, MA projected QCD, the monopole part, 
MC projected QCD and the photon part in the lattice unit.}
\end{center}
\vspace{-0.75cm}
\end{table}

\section{N-$\Delta$ mass splitting in MA projected QCD}
In this section, we study the mass spectrum of light hadrons, 
$\pi$, $\rho$, nucleon (N) and $\Delta$, in MA projected QCD. 
In particular, we focus on the N-$\Delta$ mass splitting and its origin. 
In the ordinary constituent quark model, the N-$\Delta$ mass splitting 
originates from the color-magnetic interaction \cite{RGG75}. 
As another possibility of its origin, instantons are proposed to 
contribute to the N-$\Delta$ mass splitting \cite{K85}. 
The two explanations on the origin are largely different, 
because the color-magnetic interaction appears from 
the one-gluon-exchange process and is ``perturbative'', 
while the interaction caused by instantons is nonperturbative. 

To clarify whether the main origin of the N-$\Delta$ mass splitting 
is perturbative or nonperturbative, 
we utilize MA projected QCD and the monopole part, 
where the nonperturbative nature such as 
instantons almost survives \cite{M95,SM96}
but one-gluon-exchange effects are largely reduced 
due to the Abelianization and the Hodge decomposition.
In fact, the small/large reduction of the N-$\Delta$ mass splitting 
in the monopole part is expected to indicate 
the nonperturbative/perturbative origin.

Based on this strategy, we perform the lattice QCD calculation 
with Wilson quarks 
at $\beta=6.0$ ($a \simeq 0.1{\rm fm}$) on $16^3\times 32$.
The used hopping parameter $\kappa$ and the critical hopping parameter 
$\kappa_c$ where the pion is massless are summarized in Table 2.
We find that the value of $\kappa_c$ is significantly different 
among SU(3) QCD, MA projected QCD and the monopole part. 
Then, we compare the N-$\Delta$ mass splitting among them 
in the chiral limit, where the physical situation is the same. 

\begin{table}
\label{tab2}
\begin{center}
\begin{tabular}{cccc}
\hline
\hline
& $\kappa$ & $\kappa_c$ & Gauge configuration\\
\hline
SU(3) QCD        & 0.1520, \ 0.1540, \ 0.1550           & 0.1571 & 100\\
MA projected QCD & 0.1350, \ 0.1365, \ 0.1370, \ 0.1380 & 0.1422 & 250\\
Monopole part    & 0.1260, \ 0.1270, \ 0.1275, \ 0.1282 & 0.1392 & 250\\
Photon part      & 0.1260, \ 0.1270, \ 0.1275           & --     & 100\\
\hline
\hline
\end{tabular}
\caption{The hopping parameter $\kappa$ used in the calculation and 
the critical hopping parameter $\kappa_c$, for SU(3) QCD, 
MA projected QCD, the monopole and the photon part.
The gauge configuration number is also listed.
}
\end{center}
\vspace{-0.3cm}
\end{table}

\begin{figure}
\begin{minipage}{.47\linewidth}
\centering
\includegraphics[width=6cm]{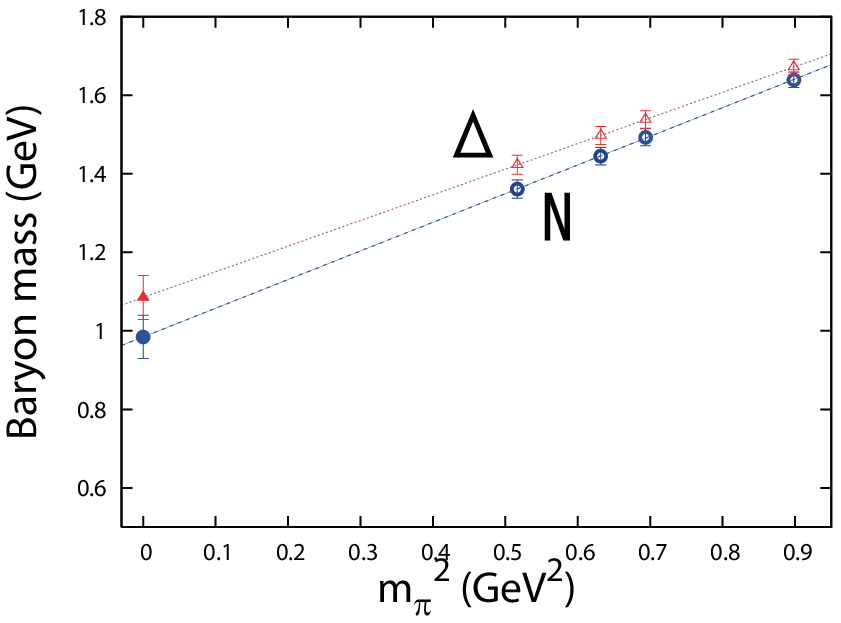}
\label{u1ph_Coul}
\caption{The mass of nucleon (N) and $\Delta$ in MA projected QCD 
plotted against the pion-mass squared, $m_\pi^2$. 
The baryon mass in chiral limit is obtained 
by standard linear chiral extrapolation.}
\end{minipage}
\hspace{0.5cm}
\begin{minipage}{.47\linewidth}
\centering
\includegraphics[width=6cm]{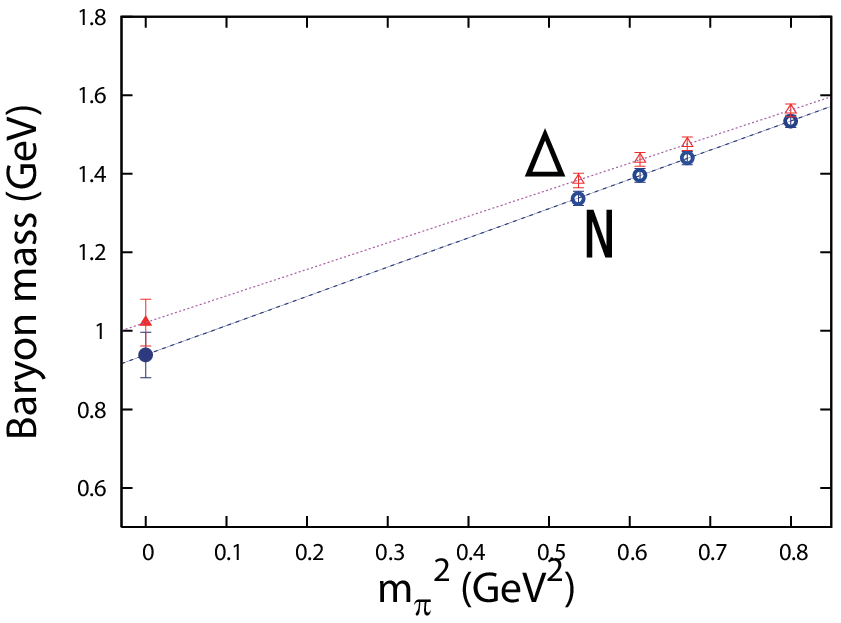}
\label{u1ph_Coul}
\caption{The mass of nucleon (N) and $\Delta$ in the monopole part 
plotted against the pion-mass squared, $m_\pi^2$. 
The baryon mass in chiral limit is obtained 
by standard linear chiral extrapolation.}
\end{minipage}
\vspace{0.3cm}
\end{figure}

\begin{table*}
\label{tab3}
\begin{center}
\begin{tabular}{cccc}
\hline
\hline
 \ \ \ \ &SU(3) QCD \ & MA projected QCD \ &Monopole part \\
\hline
N mass [GeV]          &1.058(20)  &0.985(55)  & 0.939(58) \\
$\Delta$ mass [GeV] &1.254(29)  &1.085(56)  & 1.021(60) \\
N-$\Delta$ mass splitting [GeV] & $\sim$0.20 & $\sim$0.10 & $\sim$0.08 \\
\hline
\hline
\end{tabular}
\caption{The mass of nucleon (N) and $\Delta$ 
in the chiral limit, for SU(3) QCD, MA projected QCD and 
the monopole part. 
The N-$\Delta$ mass splitting is also shown. 
The jackknife error estimation is used.}
\end{center}
\vspace{-0.5cm}
\end{table*}

Figure 7 and 8 show the mass of N and $\Delta$ 
in MA projected QCD and the monopole part, respectively. 
The numerical results in the chiral limit are summarized in Table 3. 
The N-$\Delta$ mass splitting is significantly reduced 
in MA projected QCD and the monopole part.
Such reduction can be explained by the reduction of 
the one-gluon-exchange effect, which is considered to be 
the origin of the N-$\Delta$ mass splitting in the constituent quark model.
Actually, the MA projection or the Hodge decomposition largely reduces 
one-gluon-exchange effects, which leads to the reduction of 
the Coulomb-potential part in the inter-quark potential, 
as was shown in the previous section. 
In fact, this result seems to indicate that 
the main origin of the N-$\Delta$ mass splitting is 
the one-gluon-exchange effect, which is consistent with the explanation by 
the constituent quark model.

In contrast, in the photon part, we observe the almost complete degeneracy 
between N and $\Delta$ and between $\pi$ and $\rho$ for each $\kappa$. 
The ratio of the baryon mass to the meson mass is 3 to 2, 
which is the ratio of the quark number of the hadron. 
Moreover, all the hadrons are massless in the chiral limit, 
where the pion is massless. 
These facts indicate that, in the photon part, 
the system consists of three or two quasi-free quarks, 
and no compact bound state like hadrons is created.
(We note however that the photon part is rather different from 
no interaction case with $U_\mu(s)=1$.)

\section{Summary and conclusion}
We have studied the three-quark (3Q) potential quantitatively 
in MA and MC projected QCD on quenched SU(3) lattice. 
For SU(3) QCD, MA projected QCD, the monopole part and MC projected QCD,
we have found the Y-Ansatz for the confinement potential in $V_{\rm 3Q}$ 
and the approximate equality of the string tension as 
$\sigma_{\rm 3Q}^{\rm SU(3)} \simeq\sigma_{\rm 3Q}^{\rm MA}\simeq 
\sigma_{\rm 3Q}^{\rm Mo}\simeq\sigma_{\rm 3Q}^{\rm MC}$. 
In each part, the universality of the string tension is found as 
$\sigma_{\rm 3Q} \simeq \sigma_{\rm Q\bar Q}$ between 
the 3Q and the ${\rm Q\bar Q}$ potentials. 
We have found the strong similarity of the inter-quark potential 
between monopole part and MC projected QCD. 
In contrast, almost no confinement force is found in the photon part, 
i.e., $\sigma_{\rm 3Q}^{\rm Ph}\simeq 0$. 

We have also studied the light hadron mass spectrum in MA projected QCD, 
especially focusing on the N-$\Delta$ mass splitting. 
The N-$\Delta$ mass splitting is significantly 
reduced in MA projected QCD and the monopole part, 
where the one-gluon-exchange effect or 
the Coulomb-potential part is largely reduced 
due to the Abelianization or the Hodge decomposition.
This behavior seems consistent with the constituent quark model, 
where the origin of the splitting is one-gluon exchange. 
In the photon part, the system consists of quasi-free quarks and 
compact hadrons are not created.

\end{document}